\documentclass[aps,pra,twocolumn,amsmath,amssymb,groupedaddress,longbibliography]{revtex4-1}
\usepackage{float}
\usepackage{graphicx}
\usepackage{dcolumn}
\usepackage{CJK}
\usepackage{bm}
\usepackage[dvipdfm, pdfstartview=FitH, CJKbookmarks=true, bookmarksnumbered=true, bookmarksopen=true, colorlinks=true, pdfborder=001, citecolor=blue, linkcolor=blue, linktocpage=true] {hyperref}

\def\hatd#1{\hat{#1}^\dagger}
\def\bra#1{\left\langle{#1}\right|}
\def\ket#1{\left|{#1}\right\rangle}
\def\braket#1#2{\left\langle{{#1}}\mathrel{\left|{\vphantom{{#1}{#2}}}\right.\kern-\nulldelimiterspace}{{#2}}\right\rangle}

\begin{document}


\title{Magnetic phase transitions of insulating spin-orbit coupled Bose atoms in one-dimensional optical lattices}

\author{Li Zhang$^{1,2}$}

\author{Yongguan Ke$^{1,2}$}

\author{Chaohong Lee$^{1,2}$}
\altaffiliation{Emails: lichaoh2@mail.sysu.edu.cn, chleecn@gmail.com}

\affiliation{$^{1}$Laboratory of Quantum Engineering and Quantum Metrology, School of Physics and Astronomy, Sun Yat-Sen University (Zhuhai Campus), Zhuhai 519082, China}
\affiliation{$^{2}$State Key Laboratory of Optoelectronic Materials and Technologies, Sun Yat-Sen University (Guangzhou Campus), Guangzhou 510275, China}

\date{\today}

\begin{abstract}
  We consider the insulating spin-orbit coupled Bose atoms confined within one-dimensional optical lattices and explore their ground-state magnetic phase transitions.
  Under strong interactions, the charge degrees of atoms are frozen and the system can be described by an anisotropic XXZ Heisenberg chain with Dzyaloshinskii-Moriya interaction and transverse field.
  We apply the matrix product state method to obtain low-energy states and analyze the lowest energy gaps and the ground-state magnetization and correlations.
  We find when the transverse field is absent, the ground state is a gapped ferromagnetic phase with long-range correlation in the $z$-direction if the interspin s-wave interacting strength is stronger than that of the intraspin one, otherwise it is a gapless Luttinger liquid (LL) phase with algebraic decaying correlation.
  When the transverse field is turned on, the gapless LL phase is broken, there  emerges a long-range correlated phase with ferromagnetic, antiferromagnetic or spiral order, which depends on the DM interaction strength.
  %
  %
  We believe our study provides a complete understanding of the interplay between SOC and quantum magnetism of spinor atoms in optical lattices.
\end{abstract}


\maketitle

\section{introduction\label{Sec1}}
Ultracold spinor atoms in optical lattice provide an excellent platform for simulating magnetic phase transitions in quantum Heisenberg models.
In the Mott regime, the fluctuation of charge degree of spinor atoms is suppressed and the low-energy physics can be captured by an effective spin superexchange model~\cite{Kuklov2003,Duan2003,Lee2004}, in which the spin-spin coupling can be tuned via spin-dependent s-wave scattering and lattice depth.
In recent years, there is a lot of interest in creating synthetic gauge fields and spin-orbital coupling (SOC) in ultracold system~\cite{Galitski2013,Zhai2015,Jo2018,Zhang2018}.
In optical lattice systems, SOC is engineered via dressing optical lattice with periodic Raman potentials~\cite{Liu2013,Wu2016,Song2018}, in which the SOC induces nearest-neighbor spin-flip;
and ladder-like system subject to gauge fields (the ladder legs labeled by either the internal spin states or the real-space lattices, which can be mapped to effective spin~\cite{Dario2014}) by Raman-assisted tunneling~\cite{Celi2014,Stuhl2015,Mancini2015,Atala2014} or optical clock transition~\cite{Wall2016,Livi2016,Kolkowitz2017}, in which the SOC induces on-site spin-flip.

In addition to the realistic atom-atom interaction, the SOC plays a key role in magnetic phase transitions.
It may adjust the anisotropic couplings and then lead to the so-called Dzyaloshinskii-Moriya (DM)~\cite{Dzyaloshinsky1958,Moriya1960} exchange interaction, which induces exotic magnetic phases.
In two-dimensional (2D) systems, besides normal ferromagnetic and antiferromagnetic phases, the coexistence of DM interaction and anisotropy give spiral phases, vortex crystal structure and novel Skymion~\cite{Cai2012,Cole2012,Radic2012,Gong2015,JiGuo2016}.
Classical phase diagrams have been obtained by Monte Carlo simulations~\cite{Cole2012,Gong2015}, steepest descent minimization method~\cite{Radic2012} and variational mean-field approach~\cite{JiGuo2016}.
In one-dimensional systems, the effective model reduces into an anisotropic Heisenberg chain with one-component DM interaction.
The novel structures in 2D systems is simplified to gapless spiral phase on the plane perpendicular to the DM vector, characterized by algebraic decaying correlation function, signaling a gapless Luttinger liquid (LL) phase~\cite{Zhao22014,Piraud2014,Xu2014}.
The various phases and quantum phase diagrams in 1D systems are widely studied~\cite{Peotta2014,Zhao2014,Zhao22014,Piraud2014,Xu2014} using density matrix renormalization group method.

Most previous works concentrate on nearest-neighbor spin-flip induced by SOC, where the effective models do not contain any external fields, in particular, spin-flipping fields.
An antiferromagnetic XXZ model with DM interaction and transverse field, which can be mapped into an isotropic antiferromagnetic XXX spin chain in spiral field, can be realized in a 1D fermionic ladderlike optical lattice penetrated by synthetic magnetic field~\cite{Sun2013}.
The spiral field provides a new flexible freedom to tune the phase diagram.
However, there are only two phases in this model: a gapless LL phase and a ferromagnetic phase, due to the absence of anisotropy.
One can alternately load spinor (two-component) bosons into such lattice, where the spin-dependent s-wave scattering can cause anisotropic spin-spin coupling.
Furthermore, since both fermionic~\cite{Mancini2015,Wall2016,Livi2016,Kolkowitz2017} and bosonic ladder subject to gauge field~\cite{Stuhl2015} have been realized in experiments, there is no reason to leave the bosonic case unexplored.
It is interesting to study the interplay of anisotropy, DM interaction and spin-flipping external field.

In this article, we explore the effect of SOC on the Mott insulator of two-level Bose atoms in 1D optical lattices.
Utilizing the SOC-dressed Hubbard model realized in a recent experiment~\cite{Stuhl2015}, we derive an effective spin model in the strong repulsive interaction limit: an XXZ model with DM interaction subject to external transverse field.
The strength of the anisotropy, DM interaction and transverse field of the spin model can be tuned by the parameters of the optical lattice and the SOC strength.
By employing a variational matrix product state (MPS) search~\cite{Verstraete2008,Schollwock2011,Wall2012}, we obtain the low-energy states.
Then, by calculating the lowest energy gap, and the spin-spin correlation function for the ground state (GS), we identify four typical phases characterized by long-range correlation:
(\uppercase\expandafter{\romannumeral1}) $z$-FM phase with ferromagnetic correlation along the $\hat z$ direction;
(\uppercase\expandafter{\romannumeral2}) $x$-PARA phase with only ferromagnetic correlation along the external field direction;
(\uppercase\expandafter{\romannumeral3}) $y$-AFM phase with antiferromagnetic correlation along the $\hat y$ direction; and
(\uppercase\expandafter{\romannumeral4}) $xy$-SP phase with spiral correlation on the $\hat x\hat y$ plane.
In particular, we find although the transverse field does not open the gap of the spiral phase on the $\hat x\hat y$ plane, it induces long-range correlation, distinct from the gapless LL phase found in the SOC induced nearest-neighbor spin-flip hopping model and the fermionic ladder penetrated by synthetic magnetic field.
Finally, via analyzing the order parameters, we give rich GS phase diagrams in the full parameter range.

The article structure is as follows.
In this section, we introduce the related background and our motivation.
In Sec.~\ref{Sec2}, we derive the effective Hamiltonian for our physical system.
In Sec.~\ref{Sec3}, we review the MPS method for obtaining the low-energy states.
In Sec.~\ref{Sec4}, we calculate the lowest energy gaps, two-site spin correlation function and the order parameters for determining the different phases and phase boundaries.
In the last section, we summarize and discuss our results.

\section{Strongly interacting spin-orbit coupled Bose atoms in 1D optical lattices\label{Sec2}}

Recent experiments have realized spin-orbit coupled bosons in 1D optical lattice by coupling the three internal states (pseudospins) with Raman-assisted transition~\cite{Stuhl2015}.
It is a virtual three-leg ladder pierced by magnetic flux $\phi$, with the extra dimension being the internal states of the bosons.
It can be reduced to a two-leg ladder if the second order Zeeman shift is large enough that the upper internal level can be removed~\cite{Brion2007}.
Moreover, such a bosonic ladder may also be realized by coupling metastable states of bosonic atoms by optical clock transition, which has been utilized to realize fermionic ladder in gauge field~\cite{Wall2016,Livi2016,Kolkowitz2017}.
Choosing the Landau gauge, for which the phase is accumulated by intra-leg hopping and resulting a net flux of $\phi$ through each plaquette, the single-particle Hamiltonian of this system can be written as
\begin{eqnarray}\label{Eq.Single_Ham}
  \hat H_t=&-&t\sum_{j}\left(e^{i\frac{\phi}{2}}\hatd a_{j,\uparrow}\hat a_{j+1,\uparrow}+e^{-i\frac{\phi}{2}}\hatd a_{j,\downarrow}\hat a_{j+1,\downarrow}+h.c.\right)\nonumber\\
  &-&\frac{\Omega}{2}\sum_{j}\left(\hatd a_{j,\uparrow}\hat a_{j,\downarrow}+h.c.\right),
\end{eqnarray}
where $\hatd a_{j,\sigma} (\sigma=\uparrow, \downarrow)$ creates a particle of internal state (pseudospin) $\sigma$ at site $j$.
The first term is the spin-conserved nearest-neighbor hopping with strength $te^{\pm i\frac{\phi}{2}}$, where $t$ can be tuned by the depth of the optical lattice, and the magnetic flux $\phi$ is related to the SOC momentum $k_{\mathrm{SOC}}=\phi k_L/\pi$ ($k_L$ is the lattice momentum), which is the momentum transfer of the Raman lasers and can be tuned by choosing different wavelength and/or changing the relative angle of them.
The second term describes the on-site spin-flip with strength $\frac{\Omega}{2}$, with $\Omega$ being the Rabi frequency of the Raman lasers.
Multiple particles in this optical lattice are described by the Hamiltonian $\hat H=\hat H_t+\hat H_U$, with the interaction term
\begin{equation}
\hat H_U=\frac{U}{2}\sum_{j,\sigma}\hat n_{j\sigma}\left(\hat n_{j\sigma}-1\right)+U_{\uparrow\downarrow}\sum_j\hat n_{j\uparrow}\hat n_{j\downarrow},
\end{equation}
where $\hat n_{j\sigma}=\hatd a_{j\sigma}\hat a_{j\sigma}$ is the particle number operator.
The on-site interspin and intraspin interaction strengthes are denoted as $U_{\uparrow\downarrow}$ and $U_{\sigma\sigma}$ respectively, which can be tuned by Feshbach resonance.
In the following, we set repulsive interaction as $U_{\uparrow\uparrow}=U_{\downarrow\downarrow}=U>0$ and $U_{\uparrow\downarrow}=\lambda U>0$.

We are interested in magnetic properties in the deep Mott insulator regime at half filling with strong interaction $U, U_{\uparrow\downarrow}\gg t,\Omega$.
We treat the tunneling Hamiltonian $\hat H_t$ as perturbations to the on-site interaction $\hat H_{U}$.
The GS of $\hat H_{U}$ is a Mott insulator with exactly one particle per site.
It is many-fold degenerate since the spin on every site is arbitrary.
$\hat H_{t}$ couples the manifold ground states of $\hat H_{U}$ via virtual process.
Up to second order, it gives an effective description of the low-energy physics by a spin-1/2 model.
Defining the spin operator $\mathbf{S}_j=\frac{1}{2}\hatd a_{j\alpha}\boldsymbol{\sigma}_{\alpha\beta}\hat a_{j\beta}$ with $\boldsymbol{\sigma}$ being the Pauli matrices, the effective Hamiltonian reads
\begin{eqnarray}\label{Eq.Ham_eff}
\hat H_{\mathrm{eff}}&=&J\sum_{j}\big [\cos{\phi}\left (\hat S_j^x\hat S_{j+1}^x+\hat S_j^y\hat S_{j+1}^y\right)+J_z\hat S_j^zS_{j+1}^z\nonumber\\
&+&\sin{\phi}\left (\hat S_j^x\hat S_{j+1}^y-\hat S_j^y\hat S_{j+1}^x\right)\big]-\Omega\sum_{j}\hat S_j^x,
\end{eqnarray}
where $J=-\frac{4t^2}{\lambda U}$ and $J_z=2\lambda-1$.
It is an anisotropic $XXZ$ Heisenberg model with DM interaction in transverse field.
The Heisenberg coupling and DM interaction proportional to $J$ are induced by flux dependent nearest-neighboring hopping;
The transverse field proportional to $\Omega$ is induced by on-site pseuudospin flip.
We note that the fermionic counterpart is of the same form as Eq.~\eqref{Eq.Ham_eff}, but with $J=\frac{4t^2}{U_{\uparrow\downarrow}}$ and $J_z=1$, where the magnetic transitions is a gapless LL to ferromagnetic phase transition~\cite{Sun2013}.

In the following, we study the GS properties of Eq.~\eqref{Eq.Ham_eff}.
We restrict our discussion in the regime $\phi \in [0,\pi]$, since it is the realizable range in the experiments, and Eq.~\eqref{Eq.Ham_eff} satisfies $\hat \Gamma\hat H_{\mathrm{eff}}\left(\phi, \lambda, \Omega\right)\hatd \Gamma=\hat H_{\mathrm{eff}}\left(-\phi, \lambda, \Omega\right)$ with $\hat \Gamma=\prod_{j}2\hat S_j^x$, in the theoretical perspective.
Besides, we scale the energy by $\frac{4t^2}{U}$ and define $\Omega'=\Omega/\frac{4t^2}{U}$.

\section{Matrix product state method for searching the low-energy states\label{Sec3}}

The general Hamiltonian Eq.~\eqref{Eq.Ham_eff} for arbitrary $\phi, \lambda$ and $\Omega$ are not exactly solvable.
In this section, we apply the MPS algorithm to determine the GS and low-lying excited states under open boundary condition.
It makes use of the Schmidt decomposition (SD) and treats the states and operators in the matrix product form.
By discarding the related states of small-weighted singular values, the state space is reduced in block, and states are approximated in some optimal way.
Through a variational search, the minimum-energy state in the reduced space is found.
%
%
In the following, we will introduce how to represent the states and operators in the matrix product form, and display the variational algorithm for determining the low-energy states.
%

\subsection{Schmidt decomposition}
To express the states of the system as matrix product form, one makes use of the SD.
Any pure state on a composite system $\mathcal H_A\otimes \mathcal H_B$ is read as $\ket \psi=\sum_{i_A,i_B}^{N_A,N_B}M_{i_A,i_B}\ket {i_A}\ket {i_B}$, where ${\ket {i_A}}$ and ${\ket {i_B}}$ are the bases of subsystem $A$ and $B$ with dimension $N_A$ and $N_B$ respectively.
The SD on $\ket \psi$ means it can be decomposed as $\ket \psi=\sum_{\alpha=1}^{\chi}S_{\alpha}\ket {\alpha_A}\ket {\alpha_B}$, where $\{\ket {\alpha_A}\}$ and $\{\ket{\alpha_B}\}$ are the eigenstates of the reduced density matrices $\hat\rho_A$ and $\hat\rho_B$ of the subsystem $A$ and $B$ respectively, and $S_\alpha^2$ are their shared eigenvalules, satisfying $\sum_{\alpha=1}^{\chi}S^2_\alpha=1$, with $\chi=\min(N_A,N_B)$.
The set of $\{S_\alpha\}$ are referred as Schmidt coefficients, and the number of nonzero Schmidt coefficients $\chi_s$, which meets $1\le\chi_s\le\chi$, is referred as Schmidt rank.
To relate the SD with the coefficient matrix $M$, one can expand $\{\ket{\alpha_A}\}$ and $\{\ket{\alpha_B}\}$ in the original bases as: $\ket {\alpha_A}=\sum_{i_A}U_{i_A,\alpha}\ket {i_A}$ and $\ket {\alpha_B}=\sum_{i_B}V^\dagger_{\alpha,i_B}\ket {i_B}$, with matrices $U$ and $V$ satisfying $U^\dagger U=1$ and $V^\dagger V=1$.
Defining a diagonal matrix $S$ with entries $S_\alpha$, $M$ can be expanded as $M_{i_A,i_B}=\sum_{\alpha}U_{i_A,\alpha}S_{\alpha}V^\dagger_{\alpha,i_B}$, which is just a singular value decomposition (SVD) of the $M$ matrix.
The form of the SD and the hint of the SVD on coefficient matrix $M$ provide an optimal way to approximate the state vector $\ket \psi$ with smaller spanned dimension.
That is, if the Schmidt coefficients $S_\alpha$ is arranged in a descending order: $S_1\ge S_2\ge \cdots$, and the Schmidt rank is truncated to some smaller $\tilde\chi<\chi_s$ by discarding the states with small-weighted singular value, the state
\begin{eqnarray}\label{Eq.TrunSts}
\ket {\tilde \psi}&=&\sum_\alpha^{\tilde \chi} S_\alpha\ket {\alpha_A}\ket {\alpha_B}\nonumber\\
&=&\sum_{i_A,i_B,\alpha}^{N_A,N_B,\tilde \chi} U_{i_A,\alpha}S_\alpha V^\dagger_{\alpha,i_B}\ket {i_A}\ket {i_B}
\end{eqnarray}
is the closest rank-$\tilde\chi$ approximation to $\ket \psi$ in the sense that the Frobenius norm between the coefficient matrices of these two states is minimized.
This property is the key ingredient for the feasibility of MPS algorithm, as can be seen in the following.
%

\subsection{Matrix product states}
Now, we show how to represent the quantum states of our system in matrix product form via making use of the SD or SVD.
Any pure state in 1D can be written as $\ket \psi=\sum^d_{\sigma_1\cdots\sigma_L}c_{\sigma_1\cdots\sigma_L}\ket {\sigma_1\cdots\sigma_L}$, where $\{\ket {\sigma_i}\}$ called the physical indices, are the local basis with dimension $d$, $L$ is the number of lattice sites and $c_{\sigma_1\cdots\sigma_L}$ is the complex amplitude.
This pure state can be represented as an MPS
\begin{eqnarray}
\ket \psi&=&\sum_{\vec{\sigma}}\sum_{a_0\cdots a_L}^{\chi_0\cdots\chi_L}A^{\sigma_1[1]}_{a_0,a_1}A^{\sigma_2[2]}_{a_1,a_2}\cdots A^{\sigma_L[L]}_{a_{L-1},a_L}\ket{\vec{\sigma}} \nonumber\\
&=&\sum_{\vec{\sigma}}A^{\sigma_1[1]}A^{\sigma_2[2]}\cdots A^{\sigma_L[L]}\ket {\vec{\sigma}},
\end{eqnarray}
where $\vec{\sigma}$ is shorted for $\{\sigma_i\}$, following the recursive routine:
\begin{eqnarray}\label{Eq.LOrthogonal_sts}
&&\sum_{a_l}^{\chi_l}U_{a_{l-1}\sigma_l,a_l}S_{a_l}V^\dagger_{a_l,\sigma_{l+1}\cdots\sigma_L}=\Psi_{a_{l-1}\sigma_l,\sigma_{l+1}\cdots\sigma_L},\nonumber\\
&&A_{a_{l-1},a_l}^{\sigma_l[l]}=U_{a_{l-1}\sigma_l,a_l},\nonumber\\
&&\Psi_{a_l\sigma_{l+1},\sigma_{l+2}\cdots\sigma_L}=\left(SV^\dagger\right)_{a_l,\sigma_{l+1}\cdots\sigma_L},
\end{eqnarray}
where the initial $\Psi$ is reshaped from the coefficient vector $\Psi_{a_0\sigma_1,\sigma_2\cdots\sigma_L}=c_{\sigma_1\cdots\sigma_L}$ (here $a_0=1$ is an auxiliary index), the first equality is the SVD on $\Psi$, the second equality is just the replacement of matrix $U$ by the tensor $A^{[l]}$, and the last equality is the reshaping of $SV^\dagger$ into a new matrix $\Psi$.
Each $A^{[l]}$ consists of $d$ matrices of bond dimension $\chi_{l-1}\times\chi_l$, which is determined from the SVD: $\chi_l=\min\left(d^l,d^{L-l}\right)$.
The property $U^\dagger U=I$ makes the $A^{[l]}$ satisfy the normalization relationship
\begin{eqnarray}\label{Eq.LeftNorm}
&&\sum_{\sigma_l}A^{\sigma_l[l]\dagger}A^{\sigma_l[l]}=I, \left(l<L\right)\nonumber\\
&&\sum_{\sigma_L}A^{\sigma_L[l]\dagger}A^{\sigma_L[L]}=\braket {\psi}{\psi}.
\end{eqnarray}
The MPS with all matrices satisfying this normalization condition is called left-canonical MPS.
Note that the decomposition of the complex amplitudes is not unique.
If the recursive procedure is started from the right side:
\begin{eqnarray}\label{Eq.ROrthogonal_sts}
&&\sum_{a_{l-1}}^{\chi_{l-1}}U_{\sigma_1\cdots\sigma_{l-1},a_{l-1}}S_{a_{l-1},a_{l-1}}V^\dagger_{a_{l-1},\sigma_la_l}=\Psi_{\sigma_1\cdots\sigma_{l-1},\sigma_la_l},\nonumber\\
&&B^{\sigma_l[l]}_{a_{l-1},a_l}=V^\dagger_{a_{l-1},\sigma_la_l},\nonumber\\
&&\Psi_{\sigma_1\cdots\sigma_{l-2},\sigma_{l-1}a_{l-1}}=\left(US\right)_{\sigma_1\cdots\sigma_{l-1},a_{l-1}},
\end{eqnarray}
where the initial $\Psi_{\sigma_1\cdots\sigma_{L-1},\sigma_La_L}=c_{\sigma_1\cdots\sigma_L}$ (here $a_L=1$ is the auxiliary index), the first equality is the SVD on matrix $\Psi$, the second equality is replacing $V^\dagger$ by tensor $B^{[l]}$ and the last equality is the reshaping of $US$ into a new $\Psi$, the MPS reads as
\begin{eqnarray}
\ket \psi&=&\sum_{\vec{\sigma}}\sum_{a_0\cdots a_L}^{\chi_0\cdots\chi_L}B^{\sigma_1[1]}_{a_0,a_1}B^{\sigma_2[2]}_{a_1,a_2}\cdots B^{\sigma_L[L]}_{a_{L-1},a_L}\ket{\vec{\sigma}} \nonumber\\
&=&\sum_{\vec{\sigma}}B^{\sigma_1[1]}B^{\sigma_2[2]}\cdots B^{\sigma_L[L]}\ket {\vec{\sigma}}.
\end{eqnarray}
The tensor $B^{[l]}$ consists of $d$ matrices of bond dimension $\chi_{l-1}\times\chi_l$ and satisfy the normalization condition
\begin{eqnarray}\label{Eq.RightNorm}
&&\sum_{\sigma_l}B^{\sigma_l[l]}B^{\sigma_l[l]\dagger}=I, \left(l>1\right)\nonumber\\ &&\sum_{\sigma_1}B^{\sigma_1[1]}B^{\sigma_1[1]\dagger}=\braket{\psi}{\psi},
\end{eqnarray}
making use of the fact $V^\dagger V=I$ on each SVD decomposition.
The MPS with all matrices satisfying the above normalization condition is called right-canonical MPS.

In fact, the degree of nonuniqueness is much higher: there is a gauge degree of freedom in writing the MPS.
That is, if one inserts an invertible matrix $X$ with dimension $\chi_l\times\chi_l$ and its inverse $X^{-1}$ into two adjacent MPS matrices $M^{\sigma_l[l]}$ and $M^{\sigma_{l+1}[l+1]}$ and makes the transformation $M^{\sigma_l[l]}X\to M^{\sigma_l[l]}, X^{-1}M^{\sigma_{l+1}[l+1]}\to M^{\sigma_{l+1}[l+1]}$, the MPS is invariant.
Instructively, one can specify a general MPS $\ket \psi=\sum_{\vec \sigma}M^{\sigma_1[1]}\cdots M^{\sigma_L[L]}\ket {\vec \sigma}$, by choosing a site $k$, which is called the orthogonal center, that all the matrices left and right to it are left- and right-normalized respectively.
This particular kind of MPS is called mixed-canonical MPS.
The left and right-normalization condition can be imposed by the way quite similar to the one constructing MPS from the coefficient vector.
The left-normalization condition is imposed by the recursive routine:
\begin{eqnarray}\label{Eq.LOrthogonal_mat}
&&\tilde{M}_{a_{l-1}\sigma_l,a_l}=M^{\sigma_l[l]}_{a_{l-1},a_l},\nonumber\\
&&\sum_{a'_l}U_{a_{l-1}\sigma_l,a'_l}^{[\mathrm{L}]}S_{a'_l}^{[\mathrm{L}]}V^{[\mathrm{L}]\dagger}_{a'_l,a_l}=\tilde{M}_{a_{l-1}\sigma_l,a_l},\nonumber\\
&&A^{\sigma_l[l]}_{a_{l-1},a'_l}=U_{a_{l-1}\sigma_l,a'_l}^{[\mathrm{L}]},\nonumber\\
&&M^{\sigma_{l+1}[l+1]}_{a'_l,a_{l+1}}=\sum_{a_l}\left(S^{[\mathrm{L}]}V^{[\mathrm{L}]\dagger}\right)_{a'_l,a_l}M^{\sigma_{l+1}[l+1]}_{a_l,a_{l+1}},
\end{eqnarray}
starting from $M^{[1]}$ and stoping before reaching $M^{[k]}$.
The first equality is the reshaping of tensor $M^{[l]}$ into a matrix $\tilde M$, the second equality is the SVD of $\tilde M$, the third equality is the replacement of $U^{[L]}$ by the tensor $A^{[l]}$, and the last equality is the absorption of the matrices $S^{[\mathrm{L}]}$ and $V^{[\mathrm{L}]\dagger}$ into the next $M^{[l+1]}$.
Then, the tensors $A^{[l]}$ satisfy the normalization condition Eq.~\eqref{Eq.LeftNorm}.
The bond dimension of each tensor $A^{[l]}$ is determined from the SVD: $\chi'_l=\min(d\chi_{l-1},\chi_l)$, where $\chi_l$ is the bond dimension of the original tensor $M^{[l]}$.
The right-normalization condition is enforced by the recursion routine:
\begin{eqnarray}\label{Eq.ROrthogonal_mat}
&&\tilde{M}_{a_{l-1},\sigma_la_l}=M^{\sigma_l[l]}_{a_{l-1},a_l},\nonumber\\
&&\sum_{a'_{l-1}}U_{a_{l-1},a'_{l-1}}^{[\mathrm{R}]}S_{a'_{l-1}}^{[\mathrm{R}]}V^{[\mathrm{R}]\dagger}_{a'_{l-1},\sigma_l a_l}=\tilde{M}_{a_{l-1},\sigma_la_l},\nonumber\\
&&B^{\sigma_l[l]}_{a'_{l-1},a_l}=V^{[\mathrm{R}]\dagger}_{a'_{l-1},\sigma_la_l},\nonumber\\
&&M^{\sigma_{l-1}[l-1]}_{a_{l-2},a'_{l-1}}=M^{\sigma_{l-1}[l-1]}_{a_{l-2},a_{l-1}}\left(U^{[\mathrm{R}]}S^{[\mathrm{R}]}\right)_{a_{l-1},a'_{l-1}},
\end{eqnarray}
starting from $M^{[L]}$ and stoping before reaching $M^{[k]}$.
Then, the tensors $B^{[l]}$ satisfy the normalization condition Eq.~\eqref{Eq.RightNorm}.
The bond dimension of each tensor $B^{[l]}$ is determined from the SVD: $\chi'_l=\min(d\chi_{l+1},\chi_l)$.
At last, multiply the residual $U$ and $S$ matrices resulting from the two recursion to $M^{[k]}$: $\tilde M^{\sigma_k[k]}=(S^{[\mathrm{L}]}V^{[\mathrm{L}]\dagger})M^{\sigma_k[k]}(U^{[\mathrm{R}]}S^{[\mathrm{R}]})$, and the norm square of the state is read as $\braket{\psi}{\psi}=\sum_{\sigma_k}\mathrm{Tr}\left(\tilde M^{\sigma_k[k]\dagger}\tilde M^{\sigma_k[k]}\right)$.
Definitely, if $k$ is set as $1$ or $L$, the recursion Eq.~\eqref{Eq.LOrthogonal_mat} or Eq.~\eqref{Eq.ROrthogonal_mat} gives the left- or right-canonical MPS.

The MPSs obtained in theses ways are exact, but not productive for computation, for the reason that the dimension of the matrices grows up exponentially, as can be seen from the recursion constructing the canonical MPS.
One way to make the MPSs practicable is to bound the bond dimension to some maximum $\tilde\chi$ following Eq.~\eqref{Eq.TrunSts}.
That is, in the process building a canonical MPS from a state vector or a general MPS, once the bond dimension grows above $\tilde \chi$, truncate it to $\tilde \chi$ following Eq.~\eqref{Eq.TrunSts}.
As a result, the elements in the MPS are decimated in block effectively.
This approximation is valid for GS in 1D without losing noticeable accuracy, due to two facts:
the singular value spectra decay exponentially; the bipartite entanglement of the GS obeys an area law in the gapped phase and increases as subsystem size only logarithmically near the critical point.
The MPS algorithm which introduced in the following is in fact based on such a decimation procedure.
\subsection{Matrix product operators}
The natural generation of writing states as matrix product form to operators is the matrix product operator (MPO).
A general operator $\hat O$ expressed in the local bases is $\hat O=\sum_{\vec{\sigma},\vec{\sigma}'}O_{\vec{\sigma},\vec{\sigma}'}\ket {\vec{\sigma}}\bra {\vec{\sigma}'}$.
Its matrix product form is defined as
\begin{eqnarray}\label{Eq.MPO}
\hat O&=&\sum_{\vec{\sigma},\vec{\sigma}'}\sum_{b_1\cdots b_{L-1}}^{D_1\cdots D_{L-1}}W^{\sigma_1,\sigma'_1[1]}_{1,b_1}\cdots W^{\sigma_L,\sigma'_L[L]}_{b_{L-1},1}\ket {\vec{\sigma}}\bra {\vec{\sigma}'}\nonumber\\
&=&\sum_{\{b_l\}}\hat W^{[1]}_{1,b_1}\hat W^{[2]}_{b_1,b_2}\cdots\hat W^{[L]}_{b_{L-1},1},
\end{eqnarray}
where each $\hat W^{[l]}$ can be considered as a $D_{l-1}\times D_l$ operator-valued matrix with elements $\hat W^{[l]}_{b_{l-1},b_l}=\sum_{\sigma_l,\sigma'_l}W^{\sigma_l,\sigma'_l[l]}_{b_{l-1},b_l}\ket {\sigma_l}\bra {\sigma'_l}$.
The expression of the MPO is actually a sum of matrices products, which is of the same form as a general Hamiltonian.
This makes it quite intuitive to express a 1D Hamiltonian as an MPO.
In fact, through defining some finite state automaton rules, all 1D Hamiltonian with finite-range interaction can be written as exact MPO form~\cite{Wall2012}.
In considering our model Eq.~\eqref{Eq.Ham_eff}, we can write down its MPO representation directly:
\begin{widetext}
\begin{eqnarray}\label{Eq.Ham_MPO}
\hat W^{[1]}&=&\left[\begin{array}{ccccc}
-\Omega'\hat S^x & -\frac{1}{\lambda}\left(\cos\phi\hat S^x-\sin\phi\hat S^y\right) & -\frac{1}{\lambda}\left(\cos\phi\hat S^y+\sin \phi\hat S^x\right) & -\frac{2\lambda-1}{\lambda}\hat S^z &I
\end{array}\right];\nonumber\\
\hat W^{[1<l<L]}&=&\left[\begin{array}{ccccc}
I & 0 & 0 & 0 & 0\\
\hat S^x & 0 & 0 & 0 & 0\\
\hat S^y & 0 & 0 & 0 & 0\\
\hat S^z & 0 & 0 & 0 & 0\\
-\Omega'\hat S^x & -\frac{1}{\lambda}\left(\cos\phi\hat S^x-\sin\phi\hat S^y\right) & -\frac{1}{\lambda}\left(\cos\phi\hat S^y+\sin \phi\hat S^x\right) & -\frac{2\lambda-1}{\lambda}\hat S^z &I
\end{array}\right];\ \hat W^{[L]}=\left[\begin{array}{ccccc}
I \\ \hat S^x \\ \hat S^y \\ \hat S^z \\ -\Omega'\hat S^x
\end{array}\right].\nonumber\\
\end{eqnarray}
\end{widetext}
%

\subsection{Variational ground state search}\label{SubSecGSSearch}
We now show how to obtain the GS using an MPS as a variational ansatze.
To find the optimal ground MPS $\ket\psi$ with maximum bond dimension $\chi$, we have to minimize the functional
\begin{eqnarray}\label{Eq.Energy_Functional}
\varepsilon\left[\ket \psi\right]=\bra \psi \hat H\ket\psi-E\braket{\psi}{\psi},
\end{eqnarray}
where $E$ is the Lagrangian multiplier, and $\hat H$ is in the MPO form.
This optimization problem is hard to solve at the first glace for that the variables appear as products.
Fortunately, we can get the optimal solution via an iterative algorithm: minimize the energy $\varepsilon$ with respect to the tensor $M^{[k]}$ at site $k$ with all other MPS tensors fixed, and obtain the better state lower in energy; move to the next $M^{[k+1]}$ and find the state again lower in energy; repeat sweeping through all sites until the energy is converged, and finally the minimum energy and the corresponding GS are obtained.

To minimize the energy functional $\varepsilon$ with respect to a particular $M^{[k]}$, we have to calculate $\varepsilon$ explicitly.
Suppose the ansatze MPS $\ket \psi$ is of mixed-canonical form with the orthogonal center at a chosen $k$, the overlap can be directly read as $\braket{\psi}{\psi}=\sum_{\sigma_ka_{k-1}a_k}|M^{\sigma_k[k]}_{a_{k-1},a_k}|^2$.
The expectation value of the MPO $\hat H$ in $\ket \psi$ is written as
\begin{widetext}
\begin{eqnarray}
\bra \psi \hat H\ket \psi=\sum_{\sigma_k,\sigma'_k}\sum_{a_{k-1},a_k}\sum_{a'_{k-1},a'_k}\sum_{b_{k-1},b_k} \left(L_{a_{k-1},a'_{k-1}}^{b_{k-1}[k-1]}M^{\sigma_k[k]*}_{a_{k-1},a_k}W^{\sigma_k,\sigma'_k[k]}_{b_{k-1},b_k}M^{\sigma'_k[k]}_{a'_{k-1},a'_k}R_{a_k,a'_k}^{b_k[k+1]}\right),
\end{eqnarray}
where the tensors $L$ and $R$ are the partial overlap of the Hamiltonian and the state, constructed following the recursive procedure
\begin{eqnarray}\label{Eq.Overlap_LR}
&&L_{a_l,a'_l}^{b_l[l]}=\sum_{\sigma_l,\sigma'_l}\sum_{a_{l-1},a'_{l-1}}\sum_{b_{l-1}}L_{a_{l-1},a'_{l-1}}^{b_{l-1}[l-1]}A^{\sigma_l[l]*}_{a_{l-1},a_l} W^{\sigma_l,\sigma'_l[l]}_{b_{l-1},b_l}A^{\sigma'_l[l]}_{a'_{l-1},a'_l},\nonumber\\
&&R_{a_{l-1},a'_{l-1}}^{b_{l-1}[l]}=\sum_{\sigma_l,\sigma'_l}\sum_{a_l,a'_l}\sum_{b_l}B^{\sigma_{l}[l]*}_{a_{l-1},a_l} W^{\sigma_l,\sigma'_l[l]}_{b_{l-1},b_l}B^{\sigma'_l[l]}_{a'_{l-1},a'_l}R_{a_l,a'_l}^{b_l[l+1]},
\end{eqnarray}
with the initial $L_{a_0,a'_0}^{b_0[0]}=R_{a_L,a'_L}^{b_L[L+1]}=1$.
Now take the extremum of Eq.~\eqref{Eq.Energy_Functional} with respect to $M^{\sigma_k[k]*}_{a_{k-1},a_k}$, we obtain
\begin{equation}
\sum_{\sigma'_k}\sum_{a'_{k-1},a'_k}\sum_{b_{k-1},b_k} L_{a_{k-1},a'_{k-1}}^{b_{k-1}[k-1]}W^{\sigma_k,\sigma'_k[k]}_{b_{k-1},b_k}R_{a_k,a'_k}^{b_k[k+1]}M^{\sigma'_k[k]}_{a'_{k-1},a'_k}-EM^{\sigma_k[k]}_{a_{k-1},a_k}=0.
\end{equation}
\end{widetext}
It is an eigenvalue problem $\hat H^{[k]}\ket {v^{[k]}}-E\ket {v^{[k]}}=0$ if we view $M^{[k]}$ as a vector $\ket {v^{[k]}}$ with entries $v^{[k]}_{a_{k-1}\sigma_ka_k}=M^{\sigma_k[k]}_{a_{k-1},a_k}$, and introduce the effective Hamiltonian by the reshaping
\begin{equation}\label{Eq.Hamk_var}
\hat H^{[k]}_{a_{k-1}\sigma_ka_k,a'_{k-1}\sigma'_ka'_k}=\sum_{b_{k-1},b_k}L_{a_{k-1},a'_{k-1}}^{b_{k-1}[k-1]}W^{\sigma_k,\sigma'_k[k]}_{b_{k-1},b_k}R_{a_k,a'_k}^{b_k[k+1]}.
\end{equation}
Thus the optimal solution $M^{[k]}$ at present can be obtained by solving the effective Hamiltonian $\hat H^{[k]}$ for the GS $\ket {v^{[k]}_0}$ with energy $E_0$ and reshaping $\ket {v^{[k]}_0}$ back to $M^{[k]}$, with $E_0$ being the current energy.

In summary, the iterative variational GS search algorithm is as follows:

$\left(\romannumeral1\right)$ Input. Input $\hat H$ in the MPO form, a guessed MPS $\ket \psi$ with maximum bond dimension $\chi$, and a tolerance $\varsigma$ for energy convergence.

$\left(\romannumeral2\right)$ Initialization. Transform $\ket \psi$ to the right-canonical form according to Eq.~\eqref{Eq.ROrthogonal_mat}.
Initialize the $0$th tensor $L_{a_0,a'_0}^{b_0[0]}=1$.
Construct all the right overlaps $R$ by Eq.~\eqref{Eq.Overlap_LR}.

$\left(\romannumeral3\right)$ Right sweep. Construct the effective Hamiltonian according to Eq.~\eqref{Eq.Hamk_var}, solve it for the minimum energy $E_0$ and state vector $\ket {v^{[k]}}$.
Update $M^{[k]}$ by reshaping $M^{\sigma_k[k]}_{a_{k-1},a_k}=v^{[k]}_{a_{k-1}\sigma_ka_k}$.
Left-normalize $M^{[k]}$ and move the orthogonal center to the right site $k+1$ by Eq.~\eqref{Eq.LOrthogonal_mat}.
Update the $k$th overlap $L^{[k]}$ recursively following Eq.~\eqref{Eq.Overlap_LR}.
Continue sweeping to the right until the boundary is reached.

$\left(\romannumeral4\right)$ Left sweep. Construct the effective Hamiltonian and solve it for the minimum energy $E_0$ and state vector $\ket {v^{[k]}}$.
Update $M^{[k]}$ by reshaping $M^{\sigma_k[k]}_{a_{k-1},a_k}=v^{[k]}_{a_{k-1}\sigma_ka_k}$.
Right-normalize $M^{[k]}$ and move the orthogonal center to the left site $k-1$ by Eq.~\eqref{Eq.ROrthogonal_mat}.
Update the $k$th overlap $R^{[k]}$ recursively following Eq.~\eqref{Eq.Overlap_LR}.
Continue sweeping to the left until the boundary is reached.

$\left(\romannumeral5\right)$ Repeat steps $\left(\romannumeral3\right)$ and $\left(\romannumeral4\right)$ until the convergence is achieved $\langle \hat H^2-E_0^2 \rangle<\varsigma$.

$\left(\romannumeral6\right)$ Output. Output the minimum energy $E_0$ and the MPS $\ket \psi$ which is of right-canonical form now.

Note that the state obtained in this way is not necessarily the GS, for it may get stuck in some local minimum state.
Two ways help improving such a dilemma.
The first is to prepare the initial state in the desired subspace with good quantum number.
The approached state must be the energy-minimized state in that subspace.
This is not the case for our model, for there is no explicit conserved quantity in our system.
The second way is to generalize the single site to a contiguous block during the local search and modify the algorithm accordingly, at the cost of consuming longer computational time and more computational resources.
\subsection{Variational excited states search}\label{SubSecExSSearch}

With the GS achieved, we now show how to obtain the subsequent excited states incrementally.
As the way searching the GS, the $n$th excited state ($n$ExS) is found by minimizing the energy functional $\varepsilon_n\left[\ket {\psi_n}\right]=\bra {\psi_n} \hat H\ket{\psi_n}-E_n\braket{\psi_n}{\psi_n}$, but under $n$ orthogonality constraints
\begin{equation}
 \braket {\psi_n}{\phi_m}=0,
\end{equation}
where $\{\ket {\phi_m}\}$ with $m=0, 1, \cdots n-1$ are the $n$ lower-lying eigenstates.
It makes the searching program constrained in the space orthogonal to the one formed by $\{\ket {\phi_m}\}$.
The minimization of $\varepsilon_n$ with respect to local MPS tensor $M^{[k]}$ under such constrainets is equivalent to solving the eigenvalue problem
\begin{equation}
 \left(\hat P^{[k]\dagger}\hat H^{[k]}\hat P^{[k]}\right)\ket {v^{[k]}}-E\ket {v^{[k]}}=0,
\end{equation}
where $\hat H^{[k]}$, defined by Eq.~\eqref{Eq.Hamk_var}, is the effective Hamiltonian for the variational local tensor $M^{[k]}$, and $\hat P^{[k]}$ is the project operator into the orthogonal space of the lower-lying space $\{\ket {\phi_m}\}$.

To find the projector $\hat P^{[k]}$ for every local tensor, we calculate the overlaps between the lower-lying states and the variational state explicitly: $\braket {\psi_n}{\phi_m}=\sum_{a_{k-1},\sigma_k,a_k}M_{a_{k-1},a_k}^{\sigma_k[k]*}F_{a_{k-1},a_k}^{\sigma_k[k](m)}$, with
\begin{eqnarray}\label{Eq.Overlap_var}
 F_{a_{k-1},a_k}^{\sigma_k[k](m)}=\sum_{a'_{k-1},a'_k}\mathcal{L}_{a_{k-1},a'_{k-1}}^{[k-1](m)}A^{\sigma_k[k](m)}_{a'_{k-1},a'_k}\mathcal{R}_{a_k,a'_k}^{[k+1](m)},
\end{eqnarray}
where $A^{[k](m)}$ is the $k$th MPS tensor of the $m$th lower-lying state, and the tensors $\mathcal{L}^{(m)}$ and $\mathcal{R}^{(m)}$ are the partial overlap between $\ket {\phi_m}$ and $\ket {\psi_n}$, constructed following the recursive procedure
\begin{eqnarray}\label{Eq.States_Overlap_LR}
 &&\mathcal{L}^{[l](m)}_{a_l,a'_l}=\sum_{a_{l-1},\sigma_l,a'_{l-1}}M^{\sigma_l[l]*}_{a_{l-1},a_l}\mathcal{L}^{[l-1](m)}_{a_{l-1},a'_{l-1}}A^{\sigma_l[l](m)}_{a'_{l-1},a'_l},\nonumber\\
 &&\mathcal{R}^{[l](m)}_{a_{l-1},a'_{l-1}}=\sum_{a_l,\sigma_l,a'_l}M^{\sigma_l[l]*}_{a_{l-1},a_l}\mathcal{R}^{[l+1](m)}_{a_l,a'_l}A^{\sigma_l[l](m)}_{a'_{l-1},a'_l},
\end{eqnarray}
with the initial $\mathcal{L}^{[0](m)}_{a_0,a'_0}=1$ and $\mathcal{R}^{[L+1](m)}_{a_L,a'_L}=1$.
By viewing the tensor $M^{[k]} \left(F^{[k](m)}\right)$ as vector $\ket {v^{[k]}} \left(\ket {u^{[k](m)}}\right)$ with elements $v^{[k]}_{a_{k-1}\sigma_ka_k}=M^{\sigma_k[k]}_{a_{k-1},a_k} \left( u^{[k](m)}_{a_{k-1}\sigma_ka_k}=F_{a_{k-1},a_k}^{\sigma_k[k](m)}\right )$, the projector $\hat P^{[k]}$ for the local tensor space is directly read as
\begin{equation}\label{Eq.Projector}
 \hat P^{[k]}=\hat 1-\sum_{m,m'=0}^{n-1}F^{[k](m)}\left(\mathcal N^{-1}\right)_{mm'}F^{[k](m')\dagger},
\end{equation}
where $\left(\mathcal{N}^{-1}\right)_{mm'}=\mathrm{Tr}\left(F^{[k](m)\dagger}F^{[k](m')}\right)$.

The flow of the variational $n$ExS search is the same as that for the GS, but with each step modified accordingly:

$\left(\romannumeral1\right)$ Input. Input $\hat H$ in the MPO form, the $n$ obtained lower-lying states $\{\ket {\phi_m}\}$ in the MPS form, a guessed MPS $\ket \psi$ for the $n$ExS with maximum bond dimension $\chi$, and a tolerance $\varsigma$ for energy convergence.

$\left(\romannumeral2\right)$ Initialization. Transform $\ket \psi$ to the right-canonical form according to Eq.~\eqref{Eq.ROrthogonal_mat}.
Initialize the $0$th tensors $L_{a_0,a'_0}^{b_0[0]}=1$ and $\mathcal{L}_{a_0,a'_0}^{[0](m)}=1$.
Construct all the right overlaps $R$ and $\mathcal R$ by Eq.~\eqref{Eq.Overlap_LR} and Eq.~\eqref{Eq.States_Overlap_LR} respectively.

$\left(\romannumeral3\right)$ Right sweep. Construct the effective Hamiltonian by Eq.~\eqref{Eq.Hamk_var} and the projector by Eq.~\eqref{Eq.Overlap_var} and Eq.~\eqref{Eq.Projector}. Solve the projected effective Hamiltonian for the minimum energy $E_n$ and state vector $\ket {v^{[k]}}$.
Update $M^{[k]}$ by reshaping $M^{\sigma_k[k]}_{a_{k-1},a_k}=v^{[k]}_{a_{k-1}\sigma_ka_k}$.
Left-normalize $M^{[k]}$ and move the orthogonal center to the right site $k+1$ by Eq.~\eqref{Eq.LOrthogonal_mat}.
Update the $k$th overlaps $L^{[k]}$ and $\mathcal {L}^{k}$ recursively following Eq.~\eqref{Eq.Overlap_LR} and Eq.~\eqref{Eq.States_Overlap_LR} respectively.
Continue sweeping to the right until the boundary is reached.

$\left(\romannumeral4\right)$ Left sweep. Construct the projected effective Hamiltonian and solve it for the minimum energy $E_n$ and state vector $\ket {v^{[k]}}$.
Update $M^{[k]}$ by reshaping $M^{\sigma_k[k]}_{a_{k-1},a_k}=v^{[k]}_{a_{k-1}\sigma_ka_k}$.
Right-normalize $M^{[k]}$ and move the orthogonal center to the left site $k-1$ by Eq.~\eqref{Eq.ROrthogonal_mat}.
Update the $k$th overlaps $R^{[k]}$ and $\mathcal {R}^{[k]}$ recursively following Eq.~\eqref{Eq.Overlap_LR} and Eq.~\eqref{Eq.States_Overlap_LR} respectively.
Continue sweeping to the left until the boundary is reached.

$\left(\romannumeral5\right)$ Repeat steps $\left(\romannumeral3\right)$ and $\left(\romannumeral4\right)$ until the convergence is achieved $\langle \hat H^2-E_n^2 \rangle<\varsigma$.

$\left(\romannumeral6\right)$ Output. Output the energy $E_n$ and the MPS $\ket \psi$ which is of right-canonical form now.

The variational ES search also suffers form the local minimum dilemma.
Besides, the area-law of entanglement does not apply to the bulk excited states, thus the bond dimension $\chi$ has to be increased to ensure the discarded states are of small-weighted singular values.
This limits the algorithm to be applicable only for low-energy states.
\section{Magnetic phase transitions\label{Sec4}}

In order to distinguish between different phases, we calculate the energy gaps $\Delta_{1,2}=E_{1,2}-E_0$ between the GS and the first two excited states, and expectation value of observables in the GS: the two-site correlation function $\langle\hat S_j^\alpha\hat S_l^\alpha\rangle$ and the associated structure factor $Q_\alpha(k)=\frac{1}{L}\sum_{jl}e^{ik(j-l)}\langle\hat S_j^\alpha\hat S_l^\alpha\rangle$; the order parameters, magnetization $M_\alpha=\frac{1}{L}|\sum_j\langle\hat S_j^\alpha\rangle|$, staggered magnetization $N_\alpha=\frac{1}{L}|\sum_j(-1)^j\langle\hat S_j^\alpha\rangle|$, and the spiral order $C_\alpha=\frac{1}{L}|\sum_j\langle[\boldsymbol{\vec{S}}_j\times\boldsymbol{\vec{S}}_{j+1}]^{\alpha}\rangle|$, where $\alpha=x,y,z$.

The calculation of expectation value of observables in the framework of MPS is quite feasible.
For a general observable $\hat O=\sum_{\vec \sigma,\vec \sigma'}O_{\sigma_1,\sigma'_1}^{[1]}\cdots O_{\sigma_L,\sigma'_L}^{[L]}\ket{\vec \sigma}\bra{\vec \sigma'}$, the expectation value in the state expressed as right-orthogonal MPS is directly written as
\begin{widetext}
\begin{eqnarray}
\langle \hat O\rangle=\frac{\sum\limits_{\sigma_L,\sigma'_L}O_{\sigma_L,\sigma'_L}^{[L]}M^{\sigma_L[L]\dagger}\left(\cdots\left(\sum\limits_{\sigma_2,\sigma'_2}O_{\sigma_2,\sigma'_2}^{[2]}M^{\sigma_2[2]\dagger} \left(\sum\limits_{\sigma_1,\sigma'_1}O_{\sigma_1,\sigma'_1}^{[1]}M^{\sigma_1[1]\dagger}M^{\sigma'_1[1]}\right)M^{\sigma_2[2]}\right)\cdots\right)M^{\sigma'_L[L]}}{\sum_{\sigma_1}\mathrm{Tr}\left(M^{\sigma_1[1]\dagger}M^{\sigma_1[1]}\right)},
\end{eqnarray}
\end{widetext}
where the dominate is the norm square of the state.
When reduced to single-site and two-site cases, the calculation is greatly simplified by making use of the normalization conditions.
In calculating the single-site observable $\hat S_j^{\alpha}=\sum_{\sigma_j,\sigma'_j}S^{\alpha[j]}_{\sigma_j,\sigma'_j}\ket {\sigma_j}\bra {\sigma'_j}$, it is convinent to transform the right-canonical MPS into one of mixed-canonical form with the orthogonal center at $j$.
Then the expectation value is just read as
\begin{equation}
\langle\hat S_j^\alpha\rangle=\frac{\sum_{\sigma_j,\sigma'_j}\left[S^{\alpha[j]}_{\sigma_j,\sigma'_j}\mathrm{Tr}\left(M^{\sigma_j[j]\dagger}M^{\sigma'_j[j]}\right)\right]}{\sum_{\sigma_j}\mathrm{Tr}\left(M^{\sigma_j[j]\dagger}M^{\sigma_j[j]}\right)}.
\end{equation}
In calculating the expectation value of the two-site observable $\hat S_j^\alpha\hat S_l^\beta$ (here we set $j<l$ without lose of generality), we move the orthogonal center $k$ to any site in the range $j\le k\le l$.
The evaluation of the expectation value is thus reduced to contracting the tensors in this range.
This can be done in a recursive manner: we first construct a tensor at site $j$, $G^{[j]}_{a_j,a'_j}=\sum_{\sigma_j,\sigma'_j,a_{j-1}}M^{\sigma_j[j]*}_{a_{j-1},a_j}S^{\alpha[j]}_{\sigma_j,\sigma'_j}M^{\sigma'_j[j]}_{a_{j-1},a'_j}$, then generate the next $G$ by the recursion
\begin{equation}
G^{[m]}_{a_m,a'_m}=\sum_{\substack{\sigma_m,a_{m-1}\\ \sigma'_m,a'_{m-1}}}G^{[m-1]}_{a_{m-1},a'_{m-1}}M^{\sigma_m[m]*}_{a_{m-1},a_m}O_{\sigma_m,\sigma'_m}^{[m]}M^{\sigma_m'[m]}_{a'_{m-1},a'_m},
\end{equation}
until the $l$th site is reached, where $O^{[m]}$ is the identity matrix when $m<l$ and $O^{[l]}=S^{\beta[l]}$, and finally the expectation value is calculated as $\langle \hat S^\alpha_j\hat S^\beta_l\rangle=\frac{\mathrm{Tr}\left(G^{[l]}\right)}{\mathrm{Tr}\left(M^{\sigma_k\dagger[k]}M^{\sigma_k[k]}\right)}$.

\begin{figure*}[!htbp]
\centering
  \includegraphics[width=2\columnwidth]{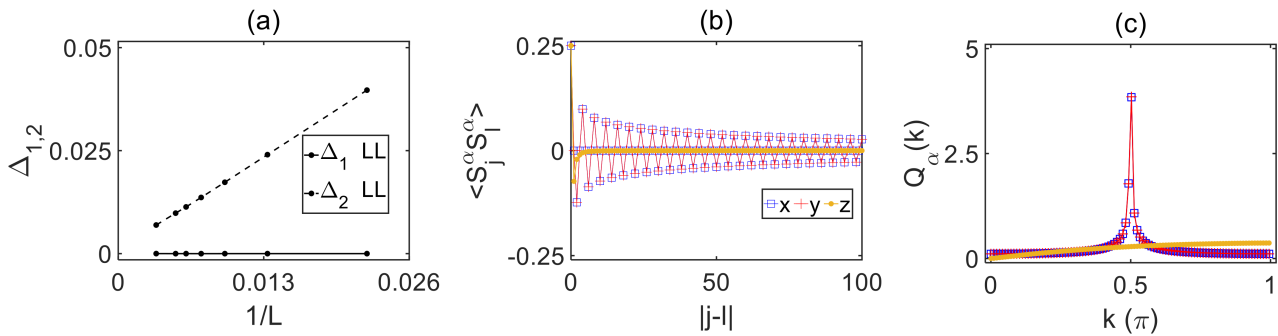}
  \caption{\label{Fig_CFunSFac_Omega0} (a) The energy gaps $\Delta_1$ (solid-dotted line) and $\Delta_2$ (dashed-dotted line) vs $1/L$; (b) Correlation function $\langle S^x_jS^x_l\rangle$ (blue squares), $\langle S^y_jS^y_l\rangle$ (red pluses) and $\langle S^z_jS^z_l\rangle$ (yellow dots) vs $|j-l|$; (c) Structure factor $Q_x\left(k\right)$ (blue squares), $Q_y \left(k\right)$ (red pluses) $Q_z\left(k\right)$ (yellow dots) vs $k$ for the LL phase of the representative parameter point $(\phi, \lambda, \Omega')=(0.5\pi, 0.75, 0)$.
  }
\end{figure*}

In our MPS calculation of the GS and the expectation value of observables, we set the system size as $L=195$, unless otherwise specified, and the tolerance for energy convergence as $\varsigma=10^{-7}$.
We randomly initialize an MPS ansatz with maximum bond dimension for several $\chi$ and searching for the GS and energy following the algorithm introduced in Sec.~\ref{SubSecGSSearch}.
We then calculate the expectation value of observables in the GS, and find the results already converges as $\chi$ approaches $16$ for $L=195$, in the sense that increasing $\chi$ does not change the results of GS energy and the expectation value of observables.
In calculating the energy gaps between the GS and the first two excited states, we fix the system size $L=45$, $75$, $105$, $135$, $165$, $195$ and $295$ respectively.
We find the maximum bond dimension has to reach $24$ for results convergence.

According to the behavior of the energy gap and correlation function, we divide the GS into four typical phases with long range correlation: (\uppercase\expandafter{\romannumeral1}) the $z$-FM phase with ferromagnetic correlation along the $\hat z$ direction; (\uppercase\expandafter{\romannumeral2}) the $x$-PARA phase only with ferromagnetic correlation along the external filed direction; (\uppercase\expandafter{\romannumeral3}) the $y$-AFM phase with antiferromagnetic correlation along the $\hat y$ direction; (\uppercase\expandafter{\romannumeral4}) the $xy$-SP phase with spiral correlation on the $\hat x\hat y$ plane.

The imposing of SOC introduces transverse field which polarizes the spin along the $\hat x$ axis, and DM interaction which causes spin to rotate on the $\hat x \hat y$ plane.
In the following, we discuss the interplay between the anisotropic Heisenberg coupling, DM interaction and transverse field.
\subsection{Interplay between anisotropy and DM interaction\label{SubSec1}}

We first study the interplay between the anisotropy and DM interaction by setting $\Omega'=0$, i.e. without Ramman-assisted tunneling.
Intuitively, the effect of the SOC should be absent and the GS properties not depend on the flux $\phi$.
To see this, we gauge away the DM interaction by performing a unitary transformation $\hat H'_{\mathrm{eff}}=\hat U\hat H_{\mathrm{eff}}\hat U^\dagger$ with $\hat U=\prod_je^{i\phi j\hat S_j^z}$, resulting in
\begin{equation}
 \hat H'_{\mathrm{eff}}=-\frac{1}{\lambda}\sum_{j}\left[\hat S_{j}^x\hat S_{j+1}^x+\hat S_{j}^y\hat S_{j+1}^y+\left(2\lambda-1\right)\hat S_j^z\hat S_{j+1}^z\right].
\end{equation}
It is a standard ferromagnetic XXZ model, whose GS has been solved exactly by the Bethe ansatz~\cite{Takahashi2005}.
As a result, when $\lambda>1$, the GS is a gapped $z$-FM phase with long-range ferromagnetic correlation in the $\hat z$ direction.
When $\lambda<1$, it is in a gapless phase with algebraic decaying correlation functions, signaling a gapless LL phase.
Regarding the original Hamiltonian, the magnetic flux $\phi$ makes the phase finer sorted.
When $\lambda>1$, the ground state is in the $z$-FM phase regardless any $\phi$.
When $\lambda<1$, the gapless LL phase transforms to a gapless ferromagnet on the $\hat x\hat y$ plane for $\phi=0$, a gapless antiferromagnet on the $\hat x\hat y$ plane for $\phi=\pi$, and a gapless spiral phase for $0<\phi<\pi$.
To illustrate the nature of the gapless spiral phase, we plot in Fig.~\ref{Fig_CFunSFac_Omega0} the finite-size scaling of the energy gaps $\Delta_{1,2}$, the correlation functions and structure factors for the representative parameter point $(\phi, \lambda)=(0.5\pi, 0.75)$.
The gapless spiral phase is characterized by zero energy gap shown in Fig.~\ref{Fig_CFunSFac_Omega0}(a) and oscillated correlation functions along the $\hat x$ and $\hat y$ direction with algebraic decaying envelop shown in Fig.~\ref{Fig_CFunSFac_Omega0}(b).
What's more, the periods of the oscillation of $\langle \hat S_j^x \hat S_l^x\rangle$ and $\langle \hat S_j^y\hat S_l^y\rangle$ are the same, as shown in Fig.~\ref{Fig_CFunSFac_Omega0}(c), where $Q_x(k)$ and $Q_y(k)$ show the same peak at $k=\phi$.
The gapless spiral phase found here is the same as that of the SOC-induced nearest-neighbor spin-flip model and the fermioic ladder, both of which belong to the LL phase.
\subsection{Interplay between anisotropy and external field}\label{SubSec2}

\begin{figure}[!htbp]
  \includegraphics[width=1\columnwidth]{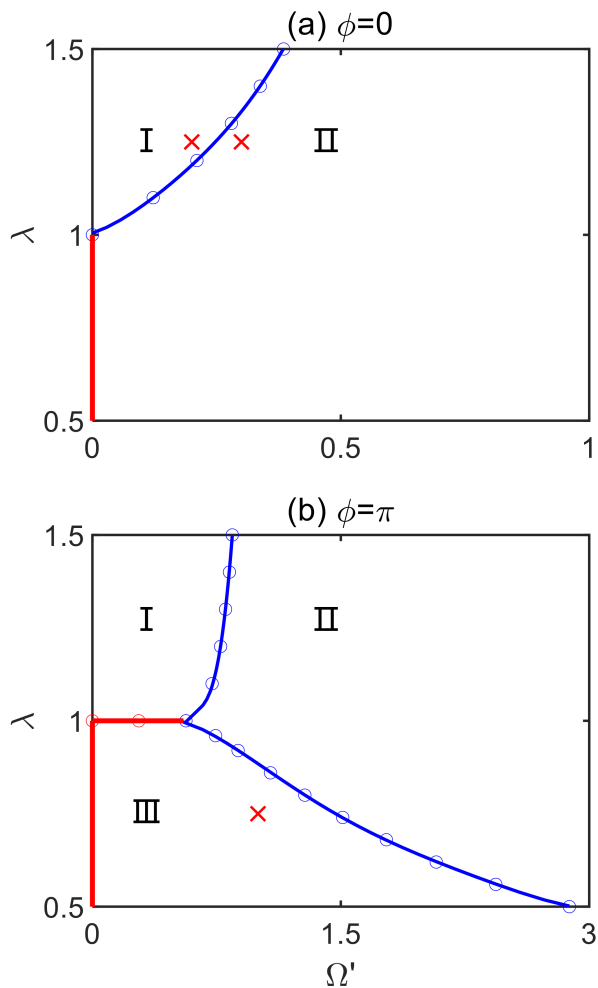}
  \caption{\label{Fig_PD_DM0} The GS phase diagrams when the DM interaction is absent by fixing (a) $\phi=0$ and (b) $\phi=\pi$.
  There are three typical phases with long range correlation: the $z$-FM phase (\uppercase\expandafter{\romannumeral1}), the $x$-PARA phase (\uppercase\expandafter{\romannumeral2}) and the $y$-AFM phase (\uppercase\expandafter{\romannumeral3}).
  The red solid lines in (a) and (b) correspond to the gapless ferromagnet and antiferromagnet respectively, both of which belong to the gapless LL phase.
  The open circles are the phase boundaries from our MPS calculation.
  The blue solid lines are fittings of the numerical data.
  The red crosses denote the parameter values at which we present the energy gaps, correlation functions and structure factors in Fig.~\ref{Fig_CFunSFac_DM0}.
  }
\end{figure}

\begin{figure*}[!htb]
  \includegraphics[width=2\columnwidth]{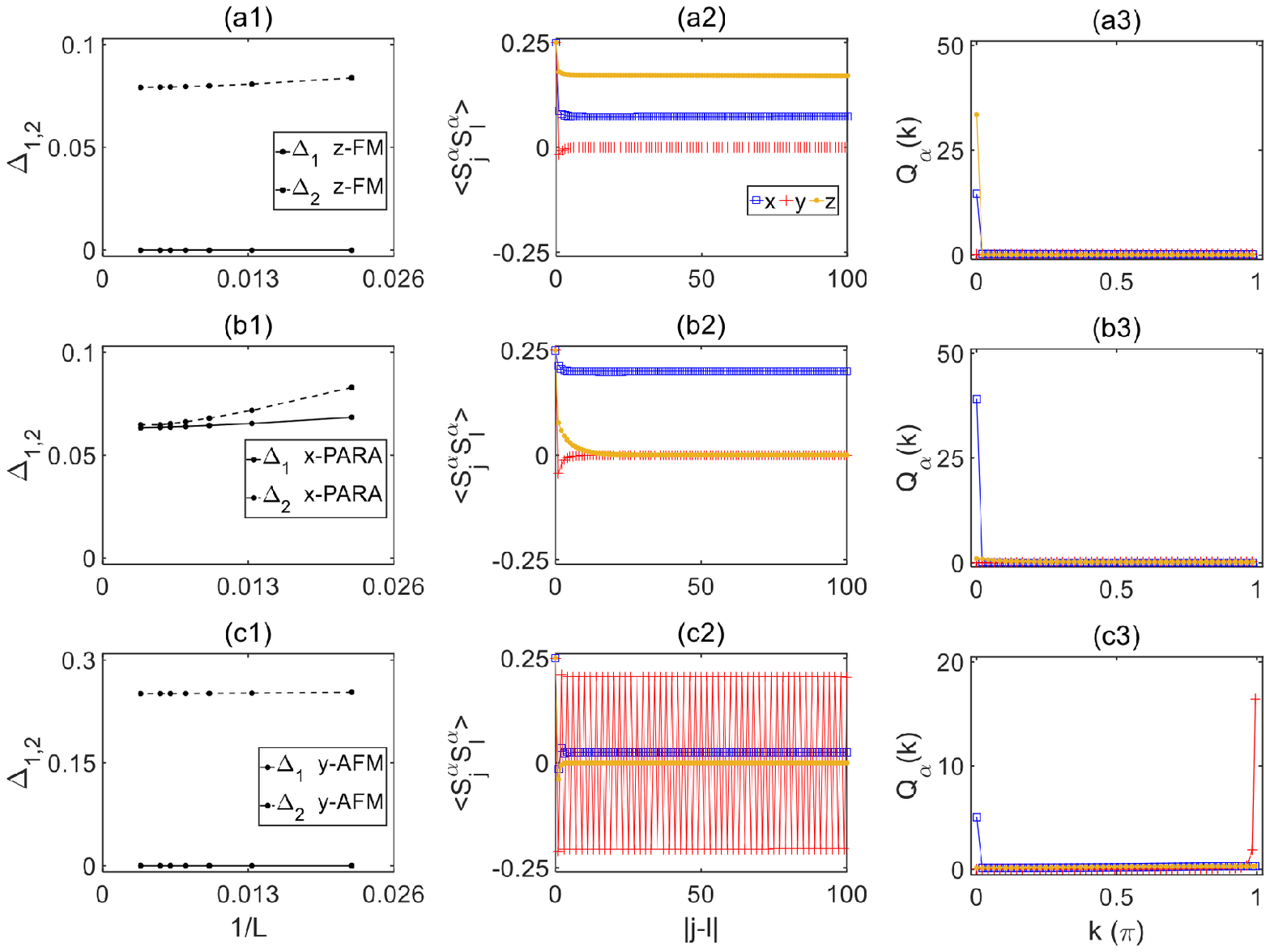}
  \caption{\label{Fig_CFunSFac_DM0} Left column: the energy gaps $\Delta_1$ (solid-dotted lines) and $\Delta_2$ (dashed-dotted lines) vs $1/L$; Middle column: correlation function $\langle S^x_jS^x_l\rangle$ (blue squares), $\langle S^y_jS^y_l\rangle$ (red pluses) and $\langle S^z_jS^z_l\rangle$ (yellow dots) vs $|j-l|$; Right column: structure factor $Q_x\left(k\right)$ (blue squares), $Q_y \left(k\right)$ (red pluses) $Q_z\left(k\right)$ (yellow dots) vs $k$.
  (a1)-(a3) The $z$-FM phase for the representative parameter point $(\phi, \lambda, \Omega')=(0, 1.25, 0.2)$.
  (b1)-(b3) The $x$-PARA phase for the representative parameter point $(\phi, \lambda, \Omega')=(0, 1.25, 0.3)$.
  (c1)-(c3) The $y$-AFM phase for the representative parameter point $(\phi, \lambda, \Omega')=(\pi, 0.75, 1)$.
  }
\end{figure*}

To discuss the interplay between anisotropic Heisenberg coupling and the external field, we set $\phi=0$ or $\phi=\pi$, i.e. zero or maximum magnetic flux, at which the DM interaction is zero.
In these two limits, the Hamiltonian is reduced to a(n) (anti)ferromagnetic XXZ model in transverse field:
\begin{eqnarray}
\hat H_{\mathrm{eff}}&=&\mp\frac{1}{\lambda}\sum_{j}\left[\hat S_j^x\hat S_{j+1}^x+\hat S_j^y\hat S_{j+1}^y\pm\left({2\lambda-1}\right)\hat S_j^zS_{j+1}^z\right] \nonumber \\
&-&\Omega'\sum_j\hat S_j^x,
\end{eqnarray}
where ferromagnetic coupling corresponds to $\phi=0$ and antiferromagnetic coupling corresponds to $\phi=\pi$.
The phase diagram of these two models are presented in Fig.~\ref{Fig_PD_DM0}, in which besides the $z$-FM (\uppercase\expandafter{\romannumeral1}) and gapless LL phase (red solid lines), additional $x$-PARA phase (\uppercase\expandafter{\romannumeral2}) and $y$-AFM phase (\uppercase\expandafter{\romannumeral3}) appear, which will be explained in the following.

We first consider the ferromagnetic case, i.e. $\phi=0$.
When $\lambda>1$, the coupling in the $\hat z$ direction dominates over that on the $\hat x\hat y$ plane.
The GS is mainly determined by the competition between the coupling in the $\hat z$ direction and the transverse field.
There is an Ising phase transition as $\Omega'$ increases: when $\Omega'$ is smaller than a threshold $\Omega'_c$, the GS exhibits long range ferromagnetic correlation along the $\hat z$ direction, so is in the $z$-FM phase.
When $\Omega'$ increases beyond $\Omega'_c$, the $z$-ferromagnetic correlation vanishes completely and only long range $x$-ferromagnetic correlation remains, thus the GS is in the so-called $x$-PARA phase.
Both of the $z$-FM and $x$-PARA phase are illustrated in Fig.~\ref{Fig_CFunSFac_DM0} (a1)-(a3) and (b1)-(b3) respectively, in which we plot the energy gaps, the correlation function $\langle S^{\alpha}_jS^{\alpha}_l\rangle$ and the related structure factors $Q_{\alpha}(k)$ for $(\lambda, \Omega')=(1.25, 0.2)$ and $(1.25, 0.3)$ respectively.
In the $z$-FM phase, the GS is twofold degenerate shown in Fig.~\ref{Fig_CFunSFac_DM0}(a1).
Both $\langle S^{z}_jS^{z}_l\rangle$ and $\langle S^{x}_jS^{x}_l\rangle$ are finite for any range and their structure factors have a peak at $k=0$, see Fig.~\ref{Fig_CFunSFac_DM0}(a2) and (a3).
In the $x$-PARA phase, there is a finite gap between the GS and the excited states, shown in Fig.~\ref{Fig_CFunSFac_DM0}(b1).
The correlation functions $\langle S^y_jS^y_l\rangle$ and $\langle S^z_jS^z_l\rangle$ decay to zero rapidly and only $\langle S^{x}_jS^{x}_l\rangle$ is finite in the long range shown in Fig.~\ref{Fig_CFunSFac_DM0}(b2), which is signatured by the smoothness of $Q_y(k)$ and $Q_z(k)$ and the sharp peak of $Q_x(k)$ at $k=0$, see Fig.~\ref{Fig_CFunSFac_DM0}(b3).
In fact, the $z$-FM$-$$x$-PARA phase transition is not unique for $\phi=0$, but also applies for arbitrary value of $\phi$ with $\lambda>1$.
In all the following discussions, we will not refer to the case with $\lambda>1$ to avoid repetition.
When $\lambda<1$, the GS is a gapless ferromagnet on the $\hat x\hat y$ plane when $\Omega'=0$, as discussed in Sec.~\ref{SubSec1}.
Nonzero $\Omega'$ breaks the degeneracy and polarizes the GS along the $\hat x$ direction immediately, meaning that there is no finite critical point between the gapless phase and the $x$-PARA phase.
We summarize the analysis above in the phase diagram for $\phi=0$ shown in Fig.~\ref{Fig_PD_DM0}(a).

We then consider the antiferromagnetic case, i.e. $\phi=\pi$.
When $\lambda=1$, via rotating every second spin around the $\hat{z}$ axis by an angle $\pi$, the Hamiltonian can be transformed into an isotropic ferromagnetic Heisenberg chain in staggered field, that is $\hat H'_{\mathrm{eff}}=-\sum_j\vec {\boldsymbol S}_j\cdot\vec {\boldsymbol S}_{j+1}-\Omega'\sum_j\left(-1\right)^j\hat S_j^x$.
The GS remains a gapless LL phase up to some critical point $\Omega'_c$, beyond which, it is a fully polarized antiferromagnetic phase~\cite{Alcaraz1995,Dmitriev2002}.
They translate to a gapless antiferromagnet and the $x$-PARA phase respectively for the original model.
When $\lambda<1$, the GS is a gapless antiferromagnet on the $\hat x\hat y$ plane when $\Omega'=0$, as discussed in Sec.~\ref{SubSec1}.
When the transverse field is turned on, its incompatibility with the antiferromagnetic coupling along the $\hat x$ direction makes the GS favor antiferromagnetic order along the $\hat y$ direction.
To manifest the nature of such a phase, we plot in Fig.~\ref{Fig_CFunSFac_DM0}(c1)-(c3) the energy gaps, the correlation functions and corresponding structure factors for a representative parameter point $(\lambda, \Omega')=(0.75, 1)$.
The GS are twofold degenerate shown in Fig.~\ref{Fig_CFunSFac_DM0}(c1); the correlation function $\langle \hat S_j^y\hat S_l^y\rangle$ displays clear long range antiferromagnetic structure shown in Fig.~\ref{Fig_CFunSFac_DM0}(c2), which is also confirmed by the sharp peak of $Q_y(k)$ at $k=\pi$ shown in Fig.~\ref{Fig_CFunSFac_DM0}(c3).
Such a antiferromagnetic structure along the $\hat y$ direction is accompanied with a finite
ferromagnetic correlation along the $\hat x$ direction, shown in Fig.~\ref{Fig_CFunSFac_DM0}(c2) and (c3), where $\langle \hat S_j^x\hat S_l^x\rangle$ is finite and $Q_x(k)$ has a smaller peak at $k=0$.
Anyway, we call such a twofold degenerate state with long range antiferomagnetic correlation in the $\hat y$ direction the $y$-AFM phase.
As $\Omega'$ increases further, the antiferromagnetic correlation breaks down and only the $x$-ferromagnetic correlation survives, and thus the GS enters the $x$-PARA phase.
There is a finite critical point separating the $y$-AFM phase and the $x$-PARA phase.
Our numerical phase diagram for $\phi=\pi$ is shown in Fig.~\ref{Fig_PD_DM0}(b), consistent with the result obtained by mean-field approach~\cite{Dmitriev2002,Dmitriev2002_2} if taking care of the definition of parameters.

\subsection{Interplay between DM interaction and external field}\label{SubSec3}

\begin{figure}[!htb]
  \includegraphics[width=1\columnwidth]{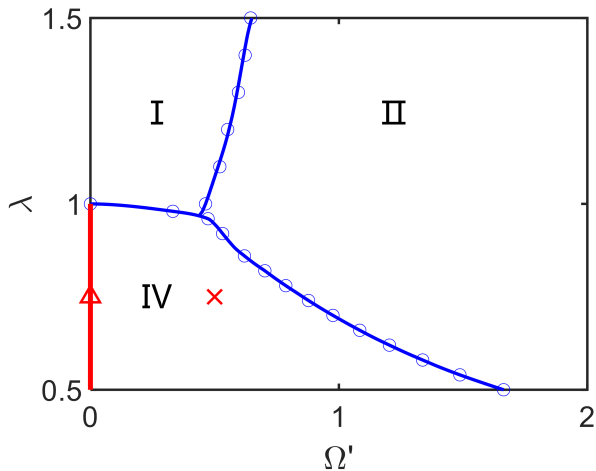}
  \caption{\label{Fig_PD_coupling0} The GS phase diagrams for $\phi=\pi/2$.
  In addition to the gapless spiral phase, which belongs to the gapless LL phase (red solid lines), the $z$-FM phase (\uppercase\expandafter{\romannumeral1}) and the $x$-FM phase (\uppercase\expandafter{\romannumeral2}), there appears the $xy$-SP phase (\uppercase\expandafter{\romannumeral4}).
  The open circles are the phase boundaries from our MPS calculation.
  The blue solid lines are fittings of the numerical data.
  The red cross and triangle at $(\lambda, \Omega')=(0.75, 0.5)$ and $(0.75, 0)$ denote the parameter values at which we present the energy gap, correlation functions and structure factors in Fig.~\ref{Fig_CFunSFac_coupling0} and Fig.~\ref{Fig_CFunSFac_Omega0} respectively.
  }
\end{figure}

\begin{figure*}[!htbp]
  \includegraphics[width=2\columnwidth]{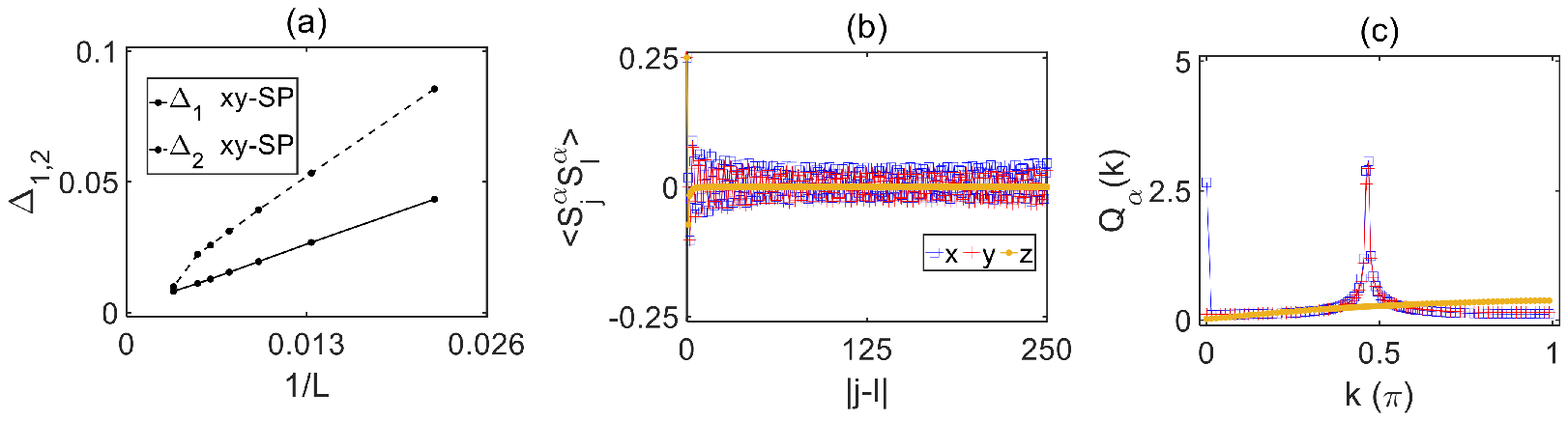}
  \caption{\label{Fig_CFunSFac_coupling0} (a) The energy gaps $\Delta_1$ (soli-dotted line) and $\Delta_2$ (dashed-dotted line) vs $1/L$; (b) Correlation function $\langle S^x_jS^x_l\rangle$ (blue squares), $\langle S^y_jS^y_l\rangle$ (red pluses) and $\langle S^z_jS^z_l\rangle$ (yellow dots) vs $|j-l|$; (c) Structure factor $Q_x\left(k\right)$ (blue squares), $Q_y \left(k\right)$ (red pluses) $Q_z\left(k\right)$ (yellow dots) vs $k$ for the $xy$-SP phase with representative parameter point $(\phi, \lambda, \Omega')=(0.5\pi, 0.75, 0.5)$.
  }
\end{figure*}

We then explore the interplay between the DM interaction and the transverse field by fixing $\phi=\pi/2$, at which the $x$-$x$ and $y$-$y$ coupling strengths are zero.
In this case, the Hamiltonian is reduced to
\begin{eqnarray}
 \hat H_{\mathrm{eff}}&=&-\frac{1}{\lambda}\sum_j\left[\left(2\lambda-1\right)\hat S_j^z\hat S_{j+1}^z+\left(\hat S_j^x\hat S_{j+1}^y-\hat S_j^y\hat S_{j+1}^x\right)\right]\nonumber\\ &-&\Omega'\sum_j\hat S_j^x.
\end{eqnarray}
When $\lambda<1$, the GS is mainly decided by the DM interaction and the transverse field.
The phase diagram when $\phi=\pi/2$ is shown in Fig.~\ref{Fig_PD_coupling0}, in which the $y$-AFM phase (\uppercase\expandafter{\romannumeral3}) is absent, and the $xy$-SP phase (\uppercase\expandafter{\romannumeral4}) emerges.

The $xy$-SP phase emerges for $0<\Omega'<\Omega'_c$.
It differs form the gapless spiral phase at $\Omega'=0$.
In Fig.~\ref{Fig_CFunSFac_coupling0}, we plot the energy gaps, correlation functions and the structure factors for the $xy$-SP phase at a representative parameter point $(\lambda,\Omega')=(0.75, 0.5)$ with system size $L=295$ and maximum bond dimension $\chi=24$.
Although the GS is maintained gapless from the excited states (at least for the first two excited states), see Fig.~\ref{Fig_CFunSFac_coupling0}(a), the correlation functions $\langle \hat S_j^x\hat S_l^x\rangle$ and $\langle \hat S_j^y\hat S_l^y\rangle$ show no decaying behaviour even in the long range, see Fig.~\ref{Fig_CFunSFac_coupling0}(b).
The periodic oscillations of $\langle \hat S_j^x\hat S_l^x\rangle$ and $\langle \hat S_j^y\hat S_l^y\rangle$ are also kept in the $xy$-SP phase, but their periods are not the same anymore, shown in Fig.~\ref{Fig_CFunSFac_coupling0}(c), where we see that besides with the shared peak with $Q_y(k)$, $Q_x(k)$ has another peak at $k=0$.
In this way, the $xy$-SP phase also distinguishes itself from the spiral phase in nearest-neighbor spin-flip model induced by SOC.
When $\Omega'$ increases beyond $\Omega'_c$, the effect of the transverse field prevails and the GS is in the $x$-PARA phase.
In this way,
The phase diagram for $\phi=\pi/2$ is shown in Fig.~\ref{Fig_PD_coupling0}.
%

\subsection{Interplay among anisotropy, DM interaction and external field}\label{SubSec4}

\begin{figure}[!htb]
  \includegraphics[width=1\columnwidth]{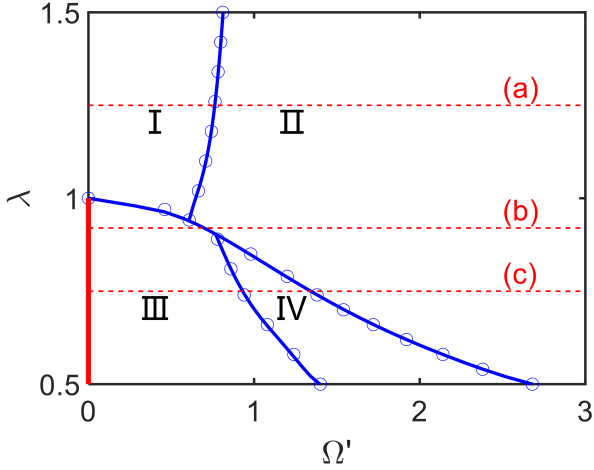}
  \caption{\label{Fig_PD_general} The GS phase diagrams for $\phi=0.8\pi$.
  There are four typical phases:
  the $z$-FM phase (\uppercase\expandafter{\romannumeral1}), the $x$-FM phase (\uppercase\expandafter{\romannumeral2}), the $y$-AFM phase (\uppercase\expandafter{\romannumeral3}) and the $xy$-SP phase (\uppercase\expandafter{\romannumeral4}).
  The red solid lines correspond to gapless LL phases.
  The open circles are the phase boundaries from our MPS calculation.
  The blue solid lines are fittings of the numerical data.
  The three red-dashed lines labeled (a), (b) and (c) correspond to $\lambda=1.25$, $0.92$ and $0.75$, along which we plot the order parameters in Fig.~\ref{Fig_OP}.
  }
\end{figure}

Finally, we discuss the general cases when the anisotropic Heisenberg coupling, the DM interaction and the transverse field are nonzero.
When $\Omega'$ increases, the system may undergo $xy$-SP$-$$x$-PARA phase transition or transit to the $y$-AFM phase before being in the $x$-PARA phase, which depends on $\phi$.
The former is the case when $0<\phi<\pi/2$, within which the $x$-$x$ and $y$-$y$ Heisenberg couplings are ferromagnetic, and only the spiral and $x$-ferromagnetic order may exist.
We have checked this argument by calculating the phase diagram by fixing $\phi<\pi/2$ and find there are only quantitative differences from Fig.~\ref{Fig_PD_coupling0}, which is not shown here.
When $\pi/2<\phi<\pi$, the $x$-$x$ and $y$-$y$ spin couplings are antiferromagnetic.
There might appear $y$-AFM phase when the antiferromagnetic coupling along with the transverse field prevails.
Thus, as $\Omega'$ increases, the system might undergo $xy$-SP$-$$y$-AFM$-$$x$-PARA phase transitions.
The phase diagram in the latter case is as in Fig.~\ref{Fig_PD_general} with fixed $\phi=0.8\pi$.

\begin{figure}[!htb]
  \includegraphics[width=1\columnwidth]{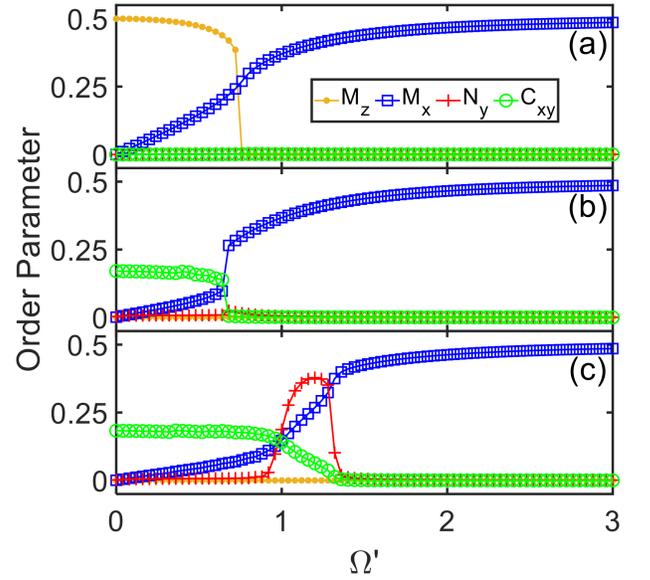}
  \caption{\label{Fig_OP} The order parameters as functions of $\Omega'$.
  The yellow lines marked by dots denote the magnetization along the $z$ direction $M_z$;
  the blue lines marked by squares denote the staggered magnetization along the $y$ direction $N_y$;
  the red lines marked by pluses denote the spiral order on the $\hat x\hat y$ plane $C_{xy}$;
  the green lines marked by circles denote the magnetization along the $x$ direction $M_x$.
  The parameters are set as $\phi=0.8\pi$ with (a) $\lambda=1.25$, (b) $0.92$ and (c) $0.75$, corresponding to the three red dashed lines in Fig.~\ref{Fig_PD_general}.
  }
\end{figure}

We now show how we determine the phase boundaries separating the four typical phases according to the behaviour of the order parameters.
Other order parameters except the ones related to the four typical phases are zero in all the parameter regime, so we do not show them further.
$M_x$ is nonzero as far as $\Omega'>0$.
In the $x$-PARA phase, all the order parameters except $M_x$ are zero.
In the $z$-FM phase, both $M_x$ and $M_z$ are non-zero.
In the $y$-AFM phase, in addition to non-zero $M_x$ and $N_y$, the incomplete $y$-AFM structure along with the extra $x$-ferromagnetic structure makes $C_{xy}$ non-zero, but suppressed compared to $M_x$ and $N_y$.
Similarly, in the $xy$-SP phase, apart from $M_x$ and $C_{xy}$, there is small amount of $N_y$.
In Fig.~\ref{Fig_PD_general}, the three red-dashed lines labeled (a), (b) and (c) correspond to $\lambda=1.25$,  $0.92$, and $0.75$ respectively.
We plot the order parameters as function of $\Omega'$ along these lines in Fig.~\ref{Fig_OP}.
In Fig.~\ref{Fig_OP}(a), there is a first-order phase transition between the $z$-FM  and the $x$-PARA phase as $\Omega'$ increases.
The transition point at $\Omega'=0.76$ can be determined by the sudden drop of $M_z$ to $0$.
In Fig.~\ref{Fig_OP}(b), with increasing $\Omega'$, there is also a first-order phase transition at $\Omega'=0.68$, separating the $xy$-SP and $x$-PARA phase.
In Fig.~\ref{Fig_OP}(c), as $\Omega'$ increases, a continuous phase transition firstly takes place at $\Omega'=0.92$, dividing the $xy$-SP and $y$-AFM phase, and then, a first-order phase transition happens at $\Omega'=1.36$ as the sudden jump of $N_y$, dividing the $y$-AFM and $x$-PARA phase.
%

\section{summary and discussions\label{Sec5}}

In summary, by employing the variational MPS method, we have studied quantum magnetic phase transitions in the deep Mott insulator of spin-orbit coupled two-level Bose atoms in 1D optical lattice.
We find that the introduction of SOC brings a DM interaction and a transverse field in the effective model, which significantly modify the magnetic structures.
When the transverse field in absent, the GS is in a gapped $z$-FM phase with ferromagnetic long range correlation in the $\hat z$ direction if the ratio of asymmetric interaction strength $\lambda>1$, and a gapless LL phase with algebraic decaying correlations if $\lambda<1$.
The gapless LL phase is classified into ferromagnet or antiferromagnet when the DM interaction is zero and spiral phase when the DM interaction in nonzero.
When the transverse field is introduced, the three kinds of gapless LL phase is broken to gapped $x$-PARA, $y$-AFM and gapless $xy$-SP phase, with finite long range correlation.
From an analysis of the order parameters, we present rich phase diagrams on the parameter spaces.
We note that the $xy$-SP phase distinguishes itself from the spiral phase in the nearest-neighbor spin-flip model induced by SOC, which is a LL phase, by the correlation function.
Thus, our study provides a complete understanding of the SOC effect on the strongly interacting atoms in optical lattice system.

Interesting extensions of our present study include the magnetic phase transition of the strongly interacting artificial bosonic three-leg ladder, whose single-particle Hamiltonian is realized in~\cite{Stuhl2015}.
The effective model will be a spin-1 chain, the GS of which would be accessible by the presented MPS algorithm.
Furthermore, in the fermionic synthetic ladder, when take the nuclear spins into consideration, the interorbital spin-exchange interaction~\cite{Cappellini2014,Scazza2014,Zhang2014} would couple individual ladders.
Thus, the model Eq.~\eqref{Eq.Single_Ham} can be generalized to multiple two-leg ladders pierced by artificial magnetic flux.
The impact of this inter-orbit coupling on the superfluid states has been thoroughly studied in~\cite{Zhou2017}.
It induces exotic vortex states on the nuclear-lattice plane, which competes with the existing phases in the decoupled ladder and leads a rich phase diagram~\cite{Zhou2017}.
How this inter-orbit coupling influences the quantum magnetism remains an open question for future research.

\begin{acknowledgments}
This work was supported by the National Natural Science Foundation of China (Grants No. 11874434 and No. 11574405).
Y.K. was partially supported by the International Postdoctoral Exchange Fellowship Program (Grant No. 20180052).
\end{acknowledgments}


\end{document}